\documentclass[journal=apchd5,manuscript=article]{achemso}

\usepackage[T1]{fontenc} % Use modern font encodings
\usepackage{graphicx}
\usepackage{fancyhdr}
\usepackage{makeidx}
\usepackage{amssymb,amsmath,amsfonts}
\usepackage{physics}
\usepackage{upquote}
\usepackage{pdfpages}
\usepackage{xcolor}
\usepackage{enumerate}
\usepackage[figurename=Figure]{caption}
\usepackage[labelfont=bf]{caption}
\newcommand{\tsup}{\textsuperscript}
\newcommand{\tsub}{\textsubscript}
\author{Medha Dandu}
\affiliation{Department of Electrical Communication Engineering, \\Indian Institute of Science, Bangalore 560012, India}
\author{Kenji Watanabe}
\affiliation{National Institute for Materials Science, \\1-1 Namiki, Tsukuba, 305-044, Japan}
\author{Takashi Taniguchi}
\affiliation{National Institute for Materials Science, \\1-1 Namiki, Tsukuba, 305-044, Japan}
\author{Ajay K. Sood}
\affiliation{Department of Physics, \\Indian Institute of Science, Bangalore 560012, India}
\author{Kausik Majumdar}
\affiliation{Department of Electrical Communication Engineering, \\Indian Institute of Science, Bangalore 560012, India}
\email{kausikm@iisc.ac.in}

\title[Spectrally tunable, large Raman enhancement from nonradiative energy transfer in van der Waals heterostructure]{Spectrally tunable, large Raman enhancement from nonradiative energy transfer in van der Waals heterostructure}

\abbreviations{1L-TMD,vdWH,NRET,PL}

\begin{document}

\begin{tocentry}
%\bibliography{ref_acsph_final}
\begin{center}
\includegraphics[scale=0.325]{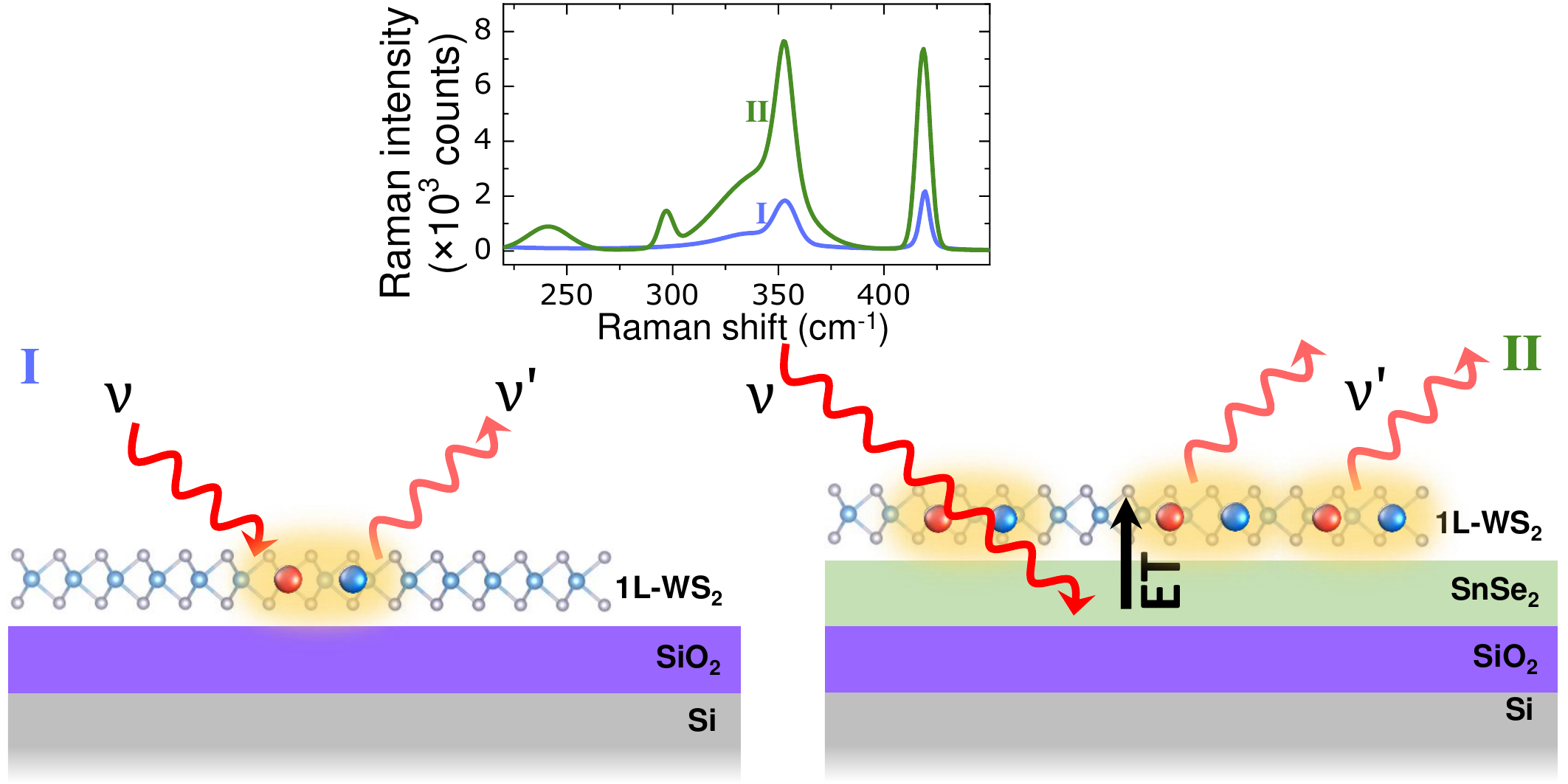}
\end{center}
\end{tocentry}

\begin{abstract}
 Raman enhancement techniques are essential for fundamental studies in light-matter interactions and find widespread application in microelectronics, bio-chemical sensing, and clinical diagnosis. Two-dimensional (2D) materials and their van der Waals heterostructures (vdWHs) are emerging rapidly as potential platforms for Raman enhancement. Here, we experimentally demonstrate a new technique of Raman enhancement driven by nonradiative energy transfer (NRET) achieving a $10$-fold enhancement in the Raman intensity in a vertical vdWH comprising of a monolayer transition metal dichalcogenide (1L-TMD) placed on a multilayer SnSe\tsub2. Consequently, several weak Raman peaks become visible which are otherwise imperceptible. We also show a strong modulation of the enhancement factor by tuning the spectral overlap between the 1L-TMD and SnSe\tsub2 through temperature variation and the results are in remarkable agreement with a Raman polarizability model capturing the effect of NRET. The observed NRET driven Raman enhancement is a novel mechanism which has not been experimentally demonstrated thus far and is distinct from conventional surface (SERS), tip (TERS) or Interference enhanced Raman scattering (IERS) mechanisms that are driven solely by charge transfer or electric field enhancement. The mechanism can also be used in synergy with plasmonic nanostructures to achieve additional selectivity and sensitivity beyond hot spot engineering for applications like molecular detection using 2D/molecular hybrids. Our results open new avenues for engineering Raman enhancement techniques coupling the advantages of uniform enhancement accessible across a wide junction area in vertical vdWHs.
\end{abstract}

\section*{Introduction}
Van der Waals heterostructures (vdWHs) offer multitude of design opportunities with diverse degrees of freedom such as choice of materials with required properties and functionalities from the vast 2D family, stacking angle, tunable interlayer interactions and heterogeneous integration with mixed dimensional systems. These design advantages coupled with the strong light-matter interactions, realization of pristine interfaces, inherent immunity to lattice mismatch, and easy prototyping render unprecedented opportunities for implementing novel device functionalities for electronics and optoelectronics\cite{song2018,zhou2018,liu2016}. These atomically sharp junctions along the vertical direction allow probing and manipulation of electronic properties in atomic length scale.
\\
\\
With closest possible physical spacing between two layers in a vdWH, Coulomb interaction enables strong interlayer dipole-dipole coupling and results in nonradiative energy transfer (NRET) from one layer (donor) to another (acceptor). The efficiency of NRET essentially depends on the physical separation and the relative orientation of the dipoles between the donor and the acceptor, the acceptor oscillator strength and the spectral overlap between the donor's emission and the acceptor's absorption\cite{guzelturk2016}. Monolayer transition metal dichalcogenides (1L-TMDs) possess large excitonic oscillator strength facilitated by strong binding energies due to confinement in the ultra-thin layer and reduced dielectric screening. The physical proximity and the in-plane orientation of the transition dipoles in each layer provide the ultimate near field coupling  generating strong NRET as proven from recent studies on photoluminescence (PL) enhancement \cite{dandu2019,gu2017,kozawa2016}.
\\
\\
Raman spectroscopy is a widely used rapid and non-invasive technique to probe light-matter interactions in vdWHs. Peak position and linewidth of Raman modes are useful probes of strain, defects and electron-phonon coupling (EPC) in 2D materials\cite{carvalho2015,chakraborty2016,das2008}. Ultra-low frequency(ULF) Raman spectroscopy reveals new Layer Breathing (LB) modes as signatures of the interlayer coupling in a vdWH\cite{lui2015,puretzky2015}.   Raman studies on vdWHs also help to probe EPC across different layers\cite{lin2019,chow2017}. Along with fundamental studies, Raman spectroscopy on 2D/molecular hybrids has recently attracted a lot of attention for analytical applications like bio-chemical sensing. 2D materials\cite{ling2014,zhang2016,miao2018} and vdWHs\cite{tan2017} have emerged as potential platforms to engineer Raman enhancement techniques such as Surface enhanced Raman scattering (SERS) substrates for strong Raman enhancement through charge transfer interaction.
\\
\\
Motivated by the strong NRET efficiency in vdWHs, here we study the prospect of Raman enhancement in a heterojunction of 1L-TMD and multilayer SnSe\tsub2 through NRET. We experimentally demonstrate NRET driven $10$-fold Raman enhancement using two different stacks, 1L-WS\tsub2/SnSe\tsub2 and 1L-MoS\tsub2/SnSe\tsub2 and also when 1L-TMD and SnSe\tsub2 are separated by a barrier layer like hBN. Interestingly, while Raman enhancement through NRET has been theoretically predicted for some systems (for example, molecules adsorbed on semiconductors through excitonic excitations in the semiconductor to the intermediate states of Raman scattering in the molecules\cite{ueba1983,dandrea1986}), to date, NRET driven Raman enhancement has not been demonstrated on any material system. Rather, NRET is commonly attributed to the quenching of the fluorescence background of molecules adsorbed on SERS metallic substrates\cite{xie2009,lee2019}.
\section*{Results and discussion}
WS\tsub2/SnSe\tsub2 (J\tsub1) and MoS\tsub2/SnSe\tsub2 (J\tsub2) samples are prepared by dry transfer method\cite{castellanos2014} on a pre-cleaned Si substrate coated with $285$ nm thick SiO\tsub2 (see \textbf{Methods}). The thickness of SnSe\tsub2 in both J\tsub1 and J\tsub2 is about 11 nm (see Figure S1 in \textbf{Supporting Information} for optical image and AFM characterization). Raman and PL measurements are carried out on these samples at $293$ K with $633$ and $532$ nm excitations. Figure 1a-b show the Raman intensity mapping images with $633$ nm excitation for J\tsub1 and J\tsub2 corresponding to $A\tsub{1g}$ mode of 1L-TMD ($419$ cm\tsup{-1} for WS\tsub2 and $405$ cm\tsup{-1} for MoS\tsub2). Both the samples exhibit strong Raman intensity enhancement on the junction area. Figure 1c-d show the representative $633$ nm Raman spectra from the isolated 1L-TMD and junction regions of both J\tsub1 and J\tsub2 depicting strong and similar enhancement of all the Raman modes on the junction, namely, $E^{1}_{2g}(\Gamma)$, $A\tsub{1g}(\Gamma)$ and $2LA(M)$ while the Si Raman peak intensity remains relatively unchanged. Note that, $E_{1g}(\Gamma)$ and $A\tsub{1g}(M)-LA(M)$ modes which are not visible on the isolated region appear on the junction\cite{zhang2015}.
\\
\\
However, with $532$ nm excitation, both J\tsub1 and J\tsub2 exhibit intensity quenching of all Raman peaks of 1L-TMD on the junction as shown in Figure 1e-f. The observation of such Raman intensity quenching with $532$ nm excitation can be correlated with the charge transfer across the junction\cite{campion1985}.  Figure 1g shows the type-I junctions formed by both MoS\tsub2 and WS\tsub2 with multilayer SnSe\tsub2. $A\tsub{1s}$ exciton peak of MoS\tsub2 and WS\tsub2 lies at 1.9 eV and 2.02 eV at $293$ K respectively while SnSe\tsub2 has a direct band gap absorption at about 2 eV\cite{li2019,evans1969,murray1973,garg197,el1992,manou1996}. SnSe\tsub2 is a degenerately doped n-type semiconductor\cite{krishna2017,aretouli2016,murali2018} with indirect band gap of 1.1 eV. On $532$ nm ($\sim2.33$ eV) excitation, carriers are excited to higher order states of 1L-TMD above $A\tsub{1s}$ as shown in Figure 1h. These excited carriers quickly relax to lower energy states available in SnSe\tsub2 at its indirect band gap through charge transfer. In vdWHs, such charge transfer occurs in the timescale of sub-ps\cite{jin2018,zhu2017}. On the other hand, the time scale of Raman process in 1L-TMD can be estimated to be tens of ps from the linewidth of Raman peaks. So, the carriers from 1L-TMD can quickly transfer to SnSe\tsub2 before scattering with a phonon mode quenching the Raman intensity on the junction. With a similar charge transfer argument, the Type-I 1L-TMD/SnSe\tsub2 junction should ideally exhibit PL intensity quenching as well. In contrast, both J\tsub1 and J\tsub2 exhibit a strong PL enhancement on the junction (Figure 1i-j) due to fast NRET process\cite{dandu2019,gu2017,kozawa2016}. Thus, despite the possibility of charge transfer induced quenching, the observed Raman enhancement in Figure 1a-d under 633 nm excitation indicates a non-trivial role of other mechanisms such as NRET and optical interference effects.
\\
\\
%\textbf{Effect of optical interference on Raman Intensity of 1L-TMD}:
The intensity of the Raman scattered light of a material is governed by the radiation characteristics of the Raman dipole and can be modified either by the Raman polarizability or by the surrounding electric field\cite{ding2017}. Any electromagnetic contribution to the change of Raman intensity ($I_{Ram}$) through electric field can be modeled by considering $I_{Ram}$ to be proportional to the product of square of amplitudes of electric field of incident ($E_{in}$) and Raman scattered ($E_{sc}$) light \cite{ramsteiner1989}, $I_{Ram}\propto |E_{in}|^{2}|E_{sc}|^{2}$. Any non-electromagnetic contribution from mechanisms such as excitation resonance, charge transfer and energy transfer, which modify the Raman polarizability, can be modeled by considering $I_{Ram}\propto |\alpha_{Ram}|^{2}$.
\\
\\
In order to decouple the extent of the roles of optical interference, excitation resonance, charge transfer and energy transfer in Raman enhancement on 1L-TMD/SnSe\tsub2, we carry out Raman measurements with $532$ and $633$ nm excitations on multiple samples of MoS\tsub2 stacked on two different materials - hBN and SnSe\tsub2, of various thicknesses (see Figure S2, S3 and S4 of \textbf{Supporting Information} for sample details). We choose hBN as a reference due to its high band gap of $\sim$ 6 eV\cite{watanabe2004,han2008} which forbids any charge or energy transfer interaction with MoS\tsub2. In order to analyze the Raman intensity modulation due to optical interference effects, we first simulate (see \textbf{Methods} for simulation details) the Raman intensity ratios for different samples from the calculations of $E_{in}$ and $E_{sc}$ on the junction ($E^{jun}_{in}$, $E^{jun}_{sc}$) and on the isolated 1L-TMD ($E^{iso}_{in}$,$E^{iso}_{sc}$) as depicted in Figure 2a.
\\
\\
To estimate the Raman intensity ratio from experimental data across different samples, we consider the ratio $\bigg(\dfrac{I^{jun}}{I^{iso}}\bigg)$ of $A\tsub{1g}$ intensity of 1L-TMD on the junction ($I^{jun}$) to that on the corresponding isolated 1L-TMD ($I^{iso}$). $A\tsub{1g}$ is chosen over $E^{1}_{2g}$ for data analysis as $A\tsub{1g}$ is relatively less sensitive to any possible strain\cite{yagmurcukardes2018}. Figure 2b shows the experimental (solid symbols) and the field simulated (open symbols) Raman intensity ratios for MoS\tsub2 as a function of hBN thickness with $532$ nm excitation, suggesting strong monotonic quenching with an increase in hBN thickness. On the other hand, the Raman intensity ratio with $633$ nm excitation exhibits a non-monotonic trend with increasing hBN thickness as shown in Figure 2c. As the trends of the experimental and simulated ratios closely follow each in both cases, we infer that Raman intensity modulation on MoS\tsub2/hBN is mainly due to interference effects. Although the field simulation matches closely with the trend of the experimental ratios, note that, for $633$ nm excitation, there is a small disparity between the two in terms of the absolute values. The origin of this disparity could be due to a possible difference in optical quality of MoS\tsub2 on hBN, doping effects\cite{wan2016}, or permanent dipole-dipole coupling between 1L-TMD and hBN due to its polar nature \cite{ling2014}.
\\
\\
On the contrary, in the MoS\tsub2/SnSe\tsub2 samples, the experimental values of Raman intensity ratio are significantly higher than the field simulation results, for both $532$ (Figure 2d) and $633$ nm (Figure 2e) excitations. In particular, for $633$ nm, the Raman intensity exhibits an overall enhancement on the junction compared to isolated MoS\tsub2 for lower SnSe\tsub2 thickness, while for $532$ nm excitation, the Raman intensity of MoS\tsub2 remains quenched on the junction for all values of SnSe\tsub2 thickness. Nonetheless, the experimental Raman intensity ratios for both excitations remain well above the field simulation predictions at any SnSe\tsub2 thickness. However, considering possible charge transfer effects due to type-I band alignment, one would rather expect a lower experimental value of Raman intensity ratio at the junction compared to the field simulation.  This contradiction, coupled with the strong light absorption properties of SnSe\tsub2 \cite{li2019}, indicates a fundamentally different mechanism, namely, NRET at play.
\\
\\
\textbf{Effect of energy transfer on Raman Intensity of 1L-TMD}:
Note that the experimental Raman intensity is $\propto |E_{in}|^{2}|E_{sc}|^{2}|\alpha_{Ram}|^{2}$, and the field simulated intensity is $\propto |E_{in}|^{2}|E_{sc}|^{2}$. By assuming that NRET only modulates the Raman polarizability $\alpha_{Ram}$ (that is, NRET and interference effects are decoupled), to quantify the Raman enhancement originating entirely from NRET mechanism, we define `NRET Raman enhancement' ($\eta^{}_{_{NRET}}$) of MoS\tsub2/SnSe\tsub2 junction as
\begin{equation}\label{eq:eta_NRET}
\eta^{}_{_{NRET}} = \dfrac{\bigg(\dfrac{I^{MoS_{2}/SnSe_{2}}}{I^{MoS_{2}}}\bigg)_{experiment}}{\bigg(\dfrac{I^{MoS_{2}/SnSe_{2}}}{I^{MoS_{2}}}\bigg)_{simulation}}
\end{equation}
and plot in red triangular symbols in the right axes of Figure 2d-e.
Such decoupling assumption is well valid in our context since the experimental data suggests that the Raman scattering by incident absorption in 1L-TMD is much weaker relative to NRET driven Raman scattering. Also, considering the possible charge transfer induced Raman intensity quenching, $\eta^{}_{_{NRET}}$ provides an estimate of the lower limit of the Raman enhancement due to NRET. The influence of other effects such as doping and strain on the Raman enhancement can be safely ignored as we do not observe any significant shift or broadening of the $E^{1}_{2g}$ and $A\tsub{1g}$ peaks on the junction relative to the isolated 1L-TMD (see Figure S5 of \textbf{Supporting Information}).
\\
\\
Figure 3a shows a schematic depiction of resonant Raman scattering driven by NRET across 1L-TMD (MoS\tsub2 or WS\tsub2) and SnSe\tsub2. The excitation resonantly excites $A\tsub{1s}$ exciton states in 1L-TMD and e-h pairs at the direct band gap in SnSe\tsub2. Because of the screening due to degenerate doping in SnSe\tsub2, it does not exhibit any excitonic features (see Figure S6 in \textbf{Supporting Information}). PL was not observed from SnSe\tsub2 even at 4K due to its indirect band gap nature\cite{li2019}. With resonant excitation, $A\tsub{1s}$ exciton in 1L-TMD undergoes exciton-phonon scattering and emits out stokes Raman scattered light\cite{dresselhaus2005}. Because of energy resonance between SnSe\tsub2 and 1L-TMD as shown in figure 1g, NRET from SnSe\tsub2 to 1L-TMD effectively enhances the exciton density and thereby enhances the intensity of Raman scattered light. Such NRET prevails even in the case of 532 nm excitation with its effect subdued due to dominant charge transfer in the type-I junction and destructive interference effects. However, the efficiency of NRET in enhancing PL and Raman is quite different as shown in Figure S5c of \textbf{Supporting Information} with PL enhancement being stronger than Raman enhancement at any SnSe\tsub2 thickness. This difference is due to the faster timescale of radiative recombination in 1L-TMD ($\sim$ sub-ps\cite{gupta2019,zhang2019,robert2016,palummo2015}) compared to the Raman processes.
\\
\\
To further validate the NRET mechanism in Raman enhancement, we introduce a spacer layer of $\sim$ $10$ nm thick hBN between MoS\tsub2 and SnSe\tsub2 as illustrated in the top panel of Figure 3b. Due to its large bandgap and thickness, hBN layer blocks any possible direct charge transfer but allows Coulomb interaction through it facilitating dipole-dipole coupling. In the bottom panel of Figure 3b, we plot the MoS\tsub2 $A\tsub{1g}$ Raman intensity ratio obtained from MoS\tsub2/hBN/SnSe\tsub2 junction ($I^{MHS}$) with MoS\tsub2/hBN ($I^{MH}$) as the control under 633 nm excitation. The experimental values are well above the field simulation results, validating NRET driven Raman enhancement arising from long-range Coulomb interaction. Raman intensity ratio from MoS\tsub2/hBN/SnSe\tsub2 junction is lower than that from MoS\tsub2/SnSe\tsub2 at similar SnSe\tsub2 thickness, because of reduced efficiency of NRET due to increased separation with hBN\cite{guzelturk2016,dandu2019}.
\\
\\
\textbf{Modulation of NRET driven Raman enhancement through spectral overlap tuning}:
The Raman enhancement across 1L-TMD/SnSe\tsub2 junction can be modulated by tuning the spectral overlap between 1L-TMD and SnSe\tsub2 as it is one of the key factors governing the NRET efficiency. Here, we achieve this spectral overlap tuning by means of shifting the exciton peak of 1L-TMD with a change in the sample temperature from $243$ to $453$ K, keeping the excitation wavelength fixed at $633$ nm ($1.96$ eV). Figure 4a shows the position of WS\tsub2 $A\tsub{1s}$ peak on isolated WS\tsub2 and junction extracted as a function of temperature using $532$ nm excitation. Isolated WS\tsub2 and junction follow a similar trend in $A\tsub{1s}$ peak position with temperature change and resonance occurs between $A\tsub{1s}$ and $633$ nm excitation around $\sim 423$ K. Similar trend is also verified from temperature dependent differential reflectance measurements (see Figure S7 in \textbf{Supporting Information}). The shift in the exciton peak of WS\tsub2 relative to the direct band gap of SnSe\tsub2 modulates the spectral overlap across 1L-TMD/SnSe\tsub2.
\\
\\
We collect  Raman spectra (at $633$ nm) from isolated WS\tsub2 and WS\tsub2/SnSe\tsub2 junction of sample J\tsub1 at different temperatures from $243$ K to $453$ K in steps of $10$ K. Figure 4b shows the plot of $A\tsub{1g}$ intensity of isolated WS\tsub2 (blue symbols) as a function of $A\tsub{1s}$ peak position where intensity goes up as exciton peak moves closer to the excitation energy with increasing temperature. To verify this trend, data in Figure 4b (blue symbols) is modeled as $I_{Ram(\omega)}\propto|\alpha_{Ram}(\omega)|^{2}$ where
\begin{equation}\label{eq:model1}
\alpha_{Ram}(\omega)= \dfrac{\lambda M_{exc}}{(\omega-\omega_{exc}-i\gamma_{exc})(\omega-\omega_{exc}-\Omega-i\gamma_{exc})}
\end{equation}
Here, $\omega$ is the excitation energy, $\omega_{exc}$ and $\gamma_{exc}$ are the energy position and the broadening of the exciton peak respectively and $\Omega$ is the $A\tsub{1g}$ phonon mode energy\cite{ueba1983,dresselhaus2005}. $\omega_{exc}$, $\gamma_{exc}$ and $\Omega$ are taken from the experimental data at the corresponding temperatures (see Figure S8 in \textbf{Supporting Information}). $\lambda$ is the electron-phonon coupling constant and $M_{exc}$ describes the matrix elements of $A\tsub{1s}$ transition in WS\tsub2 which are used as fitting constants. The peak observed in the Raman intensity profile in Figure 4b corresponds to the pole of  $\alpha_{Ram}(\omega)$ at $\omega=\omega_{exc}$ ($1.96$ eV) broadened by the influence of $\gamma_{exc}$.
\\
\\
However, Figure 4c shows the WS\tsub2 $A\tsub{1g}$ peak intensity profile on the WS\tsub2/SnSe\tsub2 junction which exhibits a completely different trend from the isolated WS\tsub2. As the isolated WS\tsub2 and the junction exhibit very similar exciton peak position (Figure 4a) at any given temperature, the difference in their Raman intensity profiles implies a change in $\alpha_{Ram}$ on the junction through interlayer coupling in the form of NRET. Any role of charge transfer in the change of $\alpha_{Ram}$ can be neglected as charge transfer process is weakly affected by the temperature. $\eta^{}_{_{NRET}}$ on the junction is modulated with temperature as highlighted in red symbols of Figure 4c along the right axis which exhibits a broad peak around $293$ K. To verify sample integrity under heating, we also sweep the temperature in the reverse direction (from $453$ to $243$ K) and similar results as forward sweep are obtained. Another WS\tsub2/SnSe\tsub2 sample, J\tsub3 (see Figure S9 of \textbf{Supporting Information}) also exhibits a very similar trend of $\eta^{}_{_{NRET}}$. Conforming to the modulation of NRET efficiency by spectral overlap tuning, PL enhancement on the junction also peaks around $293$ K (see Figure S8d in \textbf{Supporting Information}).
\\
\\
The mechanism of $\eta^{}_{_{NRET}}$ with spectral overlap tuning is schematically illustrated in Figure 4d. Left panel of Figure 4d shows representative PL spectra of WS\tsub2 on the junction at four different temperatures delineating the exciton (X) and the trion (T) peaks. The position of excitation energy ($1.96$ eV) is highlighted along the red line across these different spectra. The extreme right panel of Figure 4d represents the spectral overlap between $A\tsub{1s}$ exciton of WS\tsub2 and direct bandgap of SnSe\tsub2 at four different temperatures with $T_{1}<T_{2}<T_{3}<T_{4}$. We denote the strength of NRET between WS\tsub2 and SnSe\tsub2 by $V$ which is modulated by spectral overlap. The middle panel of Figure 4d illustrates the process of Raman scattering in WS\tsub2 on the WS\tsub2/SnSe\tsub2 junction at different temperatures. At temperature $T_{1}$ ($\approx$ $243$ K), $A\tsub{1s}$ is at higher energy above the excitation and direct gap of SnSe\tsub2. With increase in temperature to $T_{2}$ ($\approx$ $293$ K), resonance occurs between WS\tsub2 and SnSe\tsub2 resulting in strong $\eta^{}_{_{NRET}}$.  When temperature is further increased to $T_{3}$($\approx$ $423$ K) where excitation is resonant with $A\tsub{1s}$ of WS\tsub2, Raman intensity from isolated WS\tsub2 becomes maximum. However, as $A\tsub{1s}$ lowers below the band gap of SnSe\tsub2, V decreases resulting in a suppressed Raman intensity enhancement on the junction. As the NRET weakens, destructive interference in the stack starts dominating quenching the Raman intensity on the junction at $T_{3}$, despite the excitation being resonant to WS\tsub2. With further increase in temperature from $T_{3}$ to $T_{4}$ ($\approx$ $453$ K), $V$ lowers further due to further decrease in the spectral overlap. Note that, the Raman scattering is shown only from exciton states despite the overlap of trion states with the $1.96$ eV excitation. Any trion-phonon scattering is considered to be negligible due to stronger contribution of exciton-phonon scattering from a doubly resonant Raman process. This is indicated by the absence of any peak in the isolated WS\tsub2 $A\tsub{1g}$ intensity profile (Figure 4b) around trion resonance with $1.96$ eV.
\\
\\
To quantify $\eta^{}_{_{NRET}}$, we adopt a model\cite{ueba1983} by considering the photon-field interaction with the coupled Hamiltonian of the junction interacting through energy transfer. With a restriction to the first order in the displacement associated with vibrational mode of energy $\Omega$, and neglecting the dispersion of exciton states, the ratio of Raman polarizability of the junction to that of the isolated 1L-TMD is expressed as\cite{ueba1983}
\begin{equation}\label{eq:R}
R(\omega) = \bigg(\dfrac{1}{1-V^{2}g_{f}(\omega)g_{exc}^{0}(\omega)}\bigg)^{2}\bigg(1+\dfrac{M_{f}}{M_{exc}}Vg_{exc}^{0}(\omega)\bigg)^{2}
\end{equation}
$M_{exc}$ and $M_{f}$ describe the matrix elements of exciton transition and e-h transition at the energies $\omega_{exc}$ and $\omega_{f}$ in WS\tsub2 and SnSe\tsub2 respectively as illustrated in the middle panel of Figure 4d. $g_{exc}^{0}=\dfrac{1}{\omega-\omega_{exc}-i\gamma_{exc}}$ and $g_{f}=\dfrac{1}{\omega-\omega_{f}-i\gamma_{f}}$ where $\gamma_{exc}$ and $\gamma_{f}$ are the broadening of exciton and e-h transition in WS\tsub2 and SnSe\tsub2 respectively. $\eta^{}_{_{NRET}}$ shown in Figure 4c is modeled with expression of $|R(\omega)|^{2}$ from equation \ref{eq:R}. Newton-Raphson method is employed to extract $V$ by considering $\dfrac{M_{f}}{M_{exc}}$ as a constant value $M$. Due to broadband absorption in SnSe\tsub2, $M_{f}$ would not change significantly with temperature. $M_{exc}$ can also be considered to be unchanged in the experimental range of temperature as differential reflectance spectra do not show any significant change in the strength of reflection minima (see Figure S7d of \textbf{Supporting Information}). The other parameters used are $\omega=1.96$ eV, $\omega_{f}=2.02$ eV, $\gamma_{f}=0.05$ eV while $\omega_{exc}$ and $\gamma_{exc}$ are used from the experimental data. The extracted values of $V$ with two different values of $M$ shown in Figure 4e match well with the trend of $\eta^{}_{_{NRET}}$ as expected. The agreement between the experimental data and the model is remarkable in the entire temperature range. This clearly proves that the change in $V$ brought out by the spectral overlap tuning with variation in temperature modulates the Raman intensity of WS\tsub2 on its junction with SnSe\tsub2, establishing the pivotal role of NRET in Raman enhancement.
\\
\\
Such NRET driven Raman enhancement technique can also be extended to other combination of layers besides 1L-TMD and SnSe\tsub2 used here by ensuring a strongly absorbing donor material and the required spectral overlap between the donor and the acceptor for efficient energy transfer. Along with the distinct advantages of restricted in-plane momentum matching and subnanometer-scale separation between two layers in vdWHs, the thickness optimization of the layers is also essential to avoid detrimental effect of interference on the net Raman enhancement.
\\
\section*{Conclusion}
To summarize, we demonstrated nonradiative energy transfer driven $10$-fold enhancement in Raman intensity from a 1L-TMD when placed on a multilayer SnSe\tsub2. We corroborate the evidence for this mechanism by decoupling NRET from other effects of optical interference, excitation resonance and charge transfer using systematic experiments and modeling. We also demonstrated the tunability of Raman enhancement through spectral overlap modulation across 1L-TMD and SnSe\tsub2 by varying the sample temperature. Observation of such non-local energy transfer driven Raman enhancement has not been demonstrated previously on any material system and opens new ways to engineer sensing mechanisms using Raman spectroscopy in molecular systems integrated with 2D materials. This method can also be a powerful spectroscopic technique to access very weak Raman modes which are otherwise imperceptible. The technique provides uniform enhancement over a large area and thus advantageous over plasmonic nanostructure or tip enhanced based localized enhancement methods. The technique is easily integrable with other existing Raman enhancement techniques to provide additional sensitivity and functionality.

\section*{Methods}

\textbf{Heterojunction fabrication:} Different samples of vdWHs studied in this work are prepared by dry-transfer method using visco-elastic stamping with the help of a micromanipulator. The first layer, SnSe\tsub2 (from 2D Semiconductors) is either exfoliated or transferred using a PDMS substrate on a pre-cleaned surface of $285$ nm SiO\tsub2 deposited Si substrate. The thickness of SnSe\tsub2 flake is confirmed with AFM measurements. The subsequent layer of 1L-TMD (from 2D Semiconductors) is transferred from the exfoliated flake on PDMS substrate attached to a glass slide. Samples of 1L-TMD/hBN/SnSe\tsub2 are prepared with the transfer of hBN layer\cite{taniguchi2007} on to the SnSe\tsub2 leaving some portion of hBN on SiO\tsub2 before 1L-TMD transfer.  Monolayer nature of transferred 1L-TMD is confirmed from Raman and PL measurements. All the measurements at $293$ K are carried out without annealing the samples. However, measurements taken with heating the stack during or after its preparation did not show any significant change from the measurements on non-annealed samples.
\\
\textbf{Raman and PL measurements:} All the Raman and PL measurements along with Raman mapping at $293$ K on different heterostructure samples are carried out with $\times 100$ objective. $532$ and $633$ nm excitations are used at $50$ and $85$ $\mu$W of laser power respectively for exposure time of $10$ s. Temperature dependent Raman measurements are performed on a liquid nitrogen controlled temperature stage (HFS600 from Linkam Scientific) with $\times 50$ objective in nitrogen ambience.
\\
\textbf{Simulation of interference effects:} For each sample, we simulate the net incident field ($E_{in}$) at the surface of the stack without 1L-TMD on top, using the structures of the junction and the isolated regions as represented in the left panel of Figure 2a from transfer matrix reflection (TM) approach under normal incidence to calculate ($E_{in}^{jun}$) and ($E_{in}^{iso}$) respectively. To simulate the net field of scattered light ($E_{sc}$), we simulate the structures as in the right panel of Figure 2a with FDTD simulations to compute $E_{sc}^{jun}$ and $E_{sc}^{iso}$. As $A\tsub{1g}$ mode has relatively smaller Raman shift from the incident wavelength, both TM and FDTD simulations are performed at the wavelength of incident excitation. Raman intensity ratio from simulation is estimated by the product of incident field ratio, $R_{in}$ = $\frac{|E_{in}^{jun}|^2}{|E_{in}^{iso}|^2}$ calculated from TM method and scattered field ratio, $R_{sc}$ = $\frac{|E_{sc}^{jun}|^2}{|E_{sc}^{iso}|^2}$, calculated from FDTD simulations. While doing so, similar electron-phonon coupling is assumed in 1L-TMD on the junction and the isolated regions. We obtain the values of $R_{in} \sim R_{sc}$ that verifies the usual approximation $I_{Ram}\propto |E_{in}|^{4}$. Refractive index ($n$) and extinction coefficient ($k$) of Si and SiO\tsub2 exhibit wavelength dispersion and the corresponding values are taken from literature\cite{palik1998}. $k$ of hBN is assumed to be zero for $633$ nm and $532$ nm with $n$ of 1.85\cite{golla2013}. Wavelength dispersive complex refractive index values of SnSe\tsub2 are taken from the report by M. M. El-Nahass\cite{el1992}.

\section*{Associated content}
\subsection*{Supporting Information}
\begin{itemize}
	\item [] Optical images and AFM of samples used in this work
	\item [] Raman characterization from MoS\tsub2/SnSe\tsub2 samples
	\item [] Temperature dependent Raman, PL and differential reflectance spectroscopy characterization of WS\tsub2/SnSe\tsub2 sample J1.
	\item [] Raman enhancement characteristics of WS\tsub2/SnSe\tsub2 sample J3.
\end{itemize}
\section*{Author information}
\subsection*{Corresponding Author}
*E-mail: kausikm@iisc.ac.in.
\subsection*{ORCID}
\begin{itemize}
\item[] Medha Dandu: 0000-0002-5548-2882
\item[] Kenji Watanabe: 0000-0003-3701-8119
\item[] Takashi Taniguchi: 0000-0002-1467-3105
\item[] Kausik Majumdar: 0000-0002-6544-7829
\end{itemize}
\subsection*{Author contributions}
K.M. and M.D. conceived the experiment. M.D. prepared the samples and performed the
measurements. K.W. and T.T. grew the hBN crystals. M.D. and K.M. analysed and interpreted
the results with contribution from A.K.S. M.D. and K.M. wrote the paper with input from all
authors.
\subsection*{Notes}
The Authors declare no Competing Financial or Non-Financial Interests.
\begin{acknowledgement}
K. M. acknowledges the support a grant from Indian Space Research Organization (ISRO),
grants under Ramanujan Fellowship, Early Career Award, and Nano Mission from the Department
of Science and Technology (DST), Government of India, and support from MHRD,
MeitY and DST Nano Mission through NNetRA. K.W. and T.T. acknowledge support from the
Elemental Strategy Initiative conducted by the MEXT, Japan and the CREST (JPMJCR15F3),
JST.

\end{acknowledgement}


\bibliography{ref_acsph_final}
\newpage

\begin{figure*}[!hbt]
		\centering
	%\vspace{-0.5in}
	%\hspace{-0.5in}
	\includegraphics[scale=0.5] {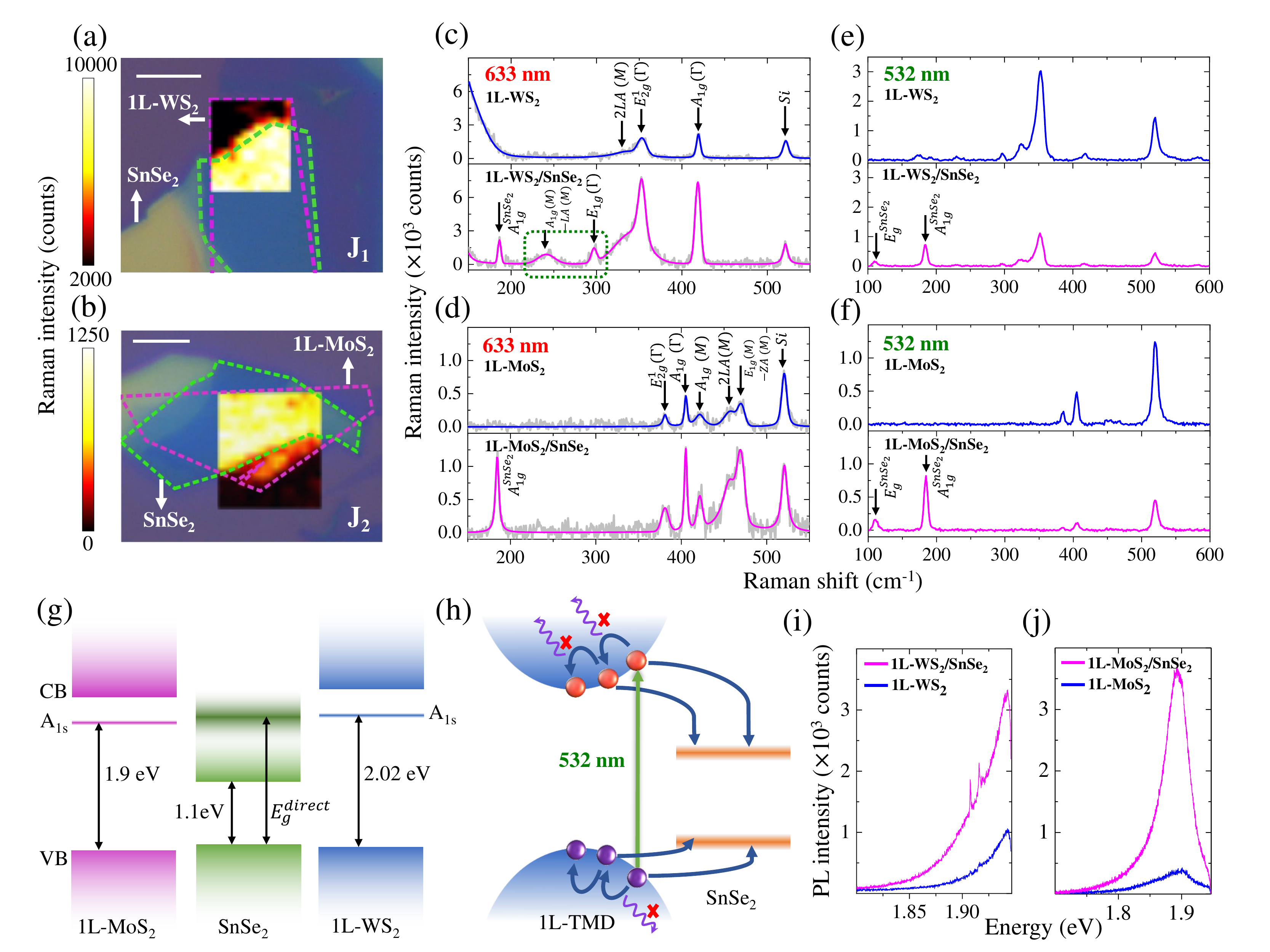}
	% \vspace{-0.1in}
	\caption{\textbf{Raman intensity enhancement on 1L-TMD/SnSe\tsub2 junction at 293 K}. (a,b) Mapping of $A_{1g}$ peak of WS\tsub2/SnSe\tsub2 (J\tsub1) and MoS\tsub2/SnSe\tsub2 (J\tsub2) respectively, acquired with 633 nm excitation. Scale bar is $5$ $\mu$m. Dashed lines highlight the boundaries of 1L-TMD (in pink) and SnSe\tsub2 (in green). (c,e) Raman spectra of isolated WS\tsub2  and WS\tsub2/SnSe\tsub2 junction with $633$ and $532$ nm excitation. (d,f) Raman spectra of isolated MoS\tsub2  and MoS\tsub2/SnSe\tsub2 junction with $633$ and $532$ nm excitation. (g) Energy band diagrams of MoS\tsub2/SnSe\tsub2 (left) and WS\tsub2/SnSe\tsub2 (right) junctions. (h) Carrier excitation in 1L-TMD with $532$ nm laser and their subsequent relaxation at 1L-TMD/SnSe\tsub2 junction which quenches the Raman intensity. (i,j) PL enhancement at 1L-TMD/SnSe\tsub2 regions of J\tsub1 and J\tsub2 with respect to isolated monolayers.}\label{fig:F1}
\end{figure*}
\pagebreak

\begin{figure*}[!hbt]
		\centering
	%\vspace{-0.5in}
	%\hspace{0.2in}
	\includegraphics[scale=0.65] {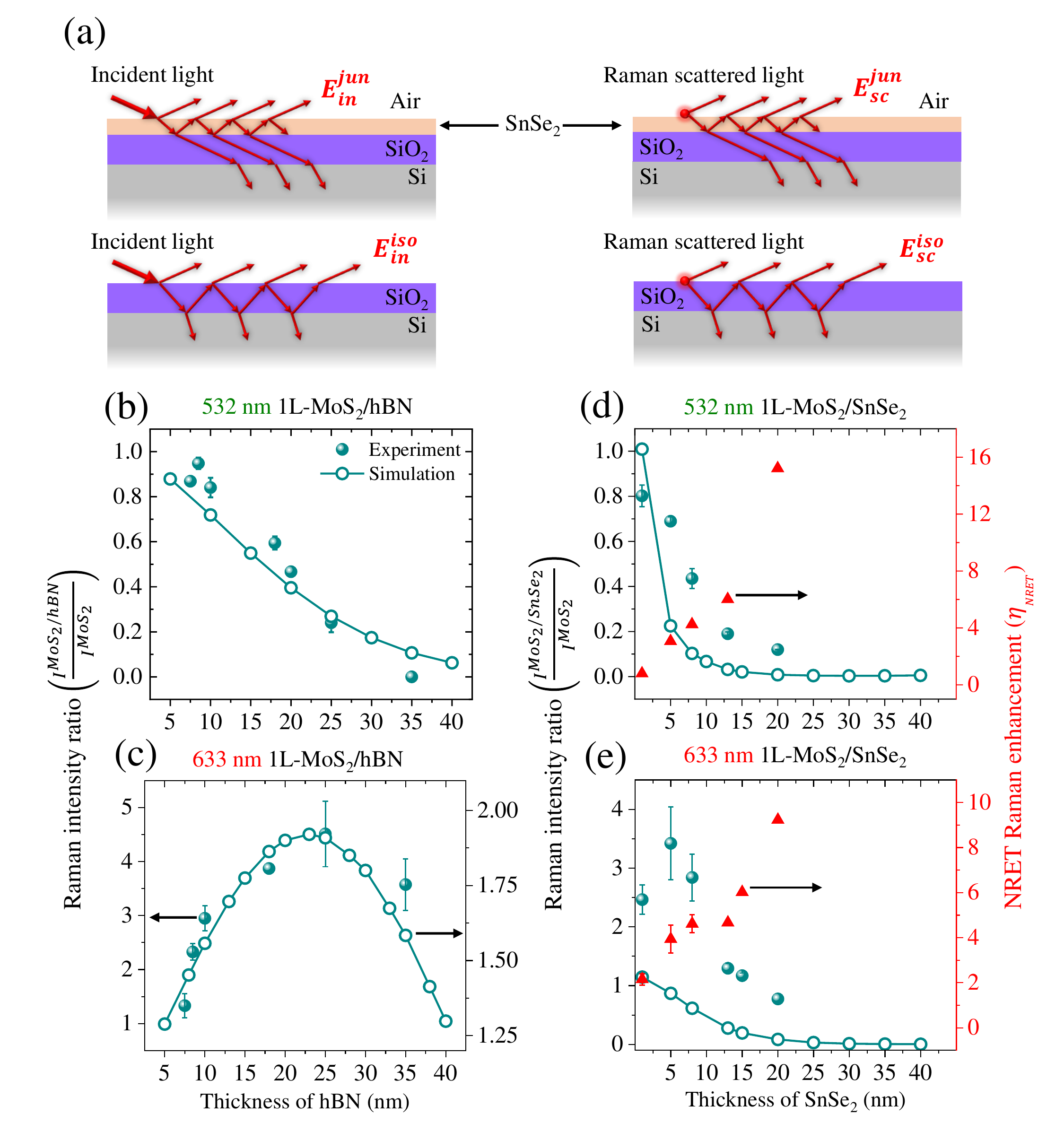}
	% \vspace{-0.1in}
	\pagebreak
	\caption{\textbf{Effect of optical interference on Raman intensity of 1L-TMD: MoS\tsub2/hBN versus MoS\tsub2/SnSe\tsub2}. (a) Schematic illustration of interference of incident light, $E_{in}$ (left panel) and Raman scattered light, $E_{sc}$ (right panel) in the heterojunction and isolated MoS\tsub2. (b,c) Experimental (solid symbols) and field simulated (open symbols) Raman intensity ratios $\big(\frac{I^{MoS_2/hBN}}{I^{MoS_2}}\big)$ with $532$ and $633$ nm excitations. (d,e) Experimental and field simulated Raman intensity ratios $\big(\frac{I^{MoS_2/SnSe_2}}{I^{MoS_2}}\big)$ with $532$ and $633$ nm excitations (left axes). Right axes in (d,e) indicate the NRET driven Raman enhancement ($\eta^{}_{_{NRET}}$) obtained by dividing the experimental ratio by the field simulation ratio (Equation \ref{eq:eta_NRET}). }\label{fig:F2}
\end{figure*}
\pagebreak
\begin{figure*}[!hbt]
		\centering
	%\vspace{-1.5in}
	%\hspace{-0.75 in}
	\includegraphics[scale=0.5] {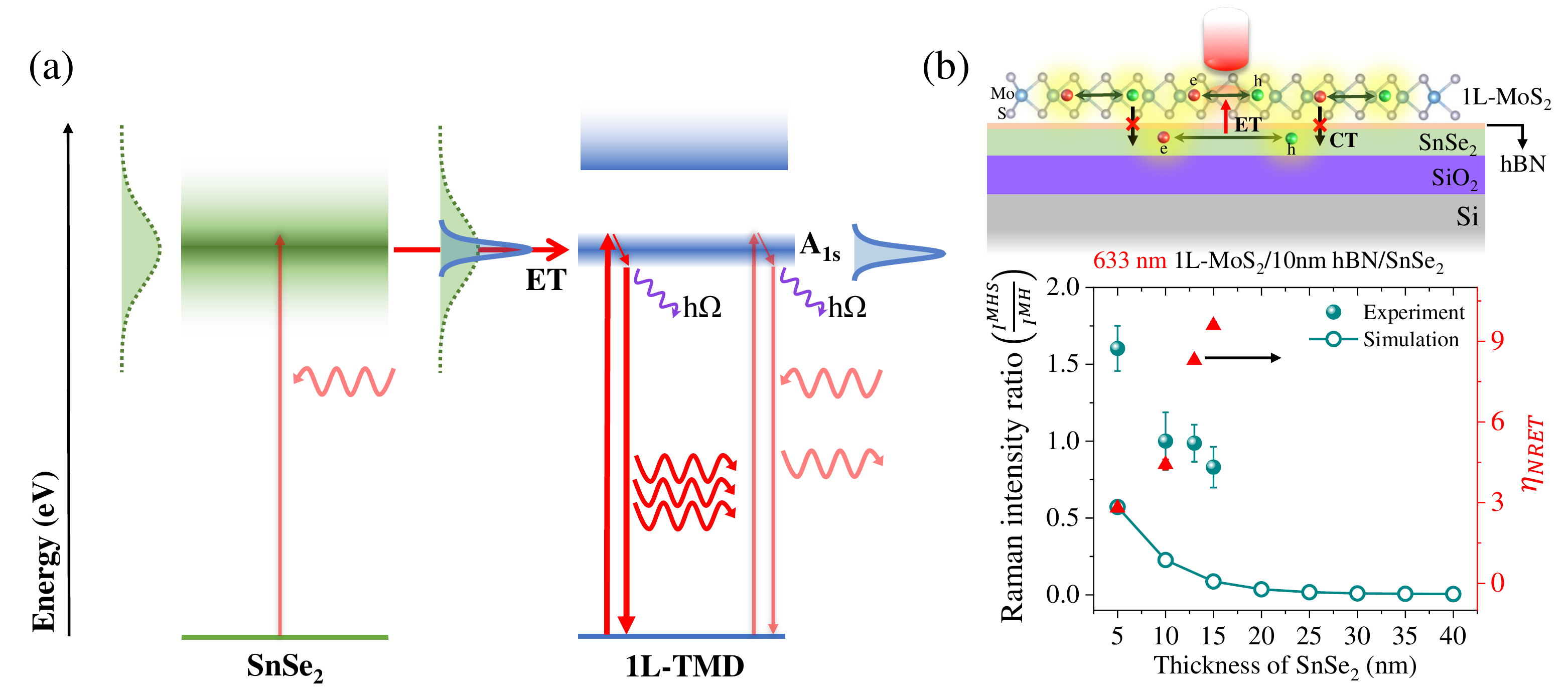}
	% \vspace{-0.1in}
	\caption{\textbf{NRET driven Raman enhancement in 1L-TMD/SnSe\tsub2}. (a) Pictorial representation of transitions in 1L-TMD at the junction following resonant Raman excitation. NRET from SnSe\tsub2 to 1L-TMD enhances the exciton population at the $A_{1s}$ state. (b) Top panel - Schematic of interlayer coupling of transitions in the MoS\tsub2/$10$ nm hBN/SnSe\tsub2 stack. $10$ nm hBN hinders charge transfer but allows NRET through Coulomb interaction. Bottom panel - Experimental and simulated Raman intensity ratios $\frac{I^{MHS}}{I^{MH}}$ under $633$ nm excitation (left axis). MHS denotes MoS\tsub2/hBN/SnSe\tsub2 and MH denotes MoS\tsub2/SnSe\tsub2. Corresponding $\eta^{}_{_{NRET}}$ is shown in the right axis.
	}\label{fig:F3}
\end{figure*}
\pagebreak
\begin{figure*}[!h]
		\centering
	%\vspace{1.5in}
	%\hspace{-1in}
	\includegraphics[scale=0.4] {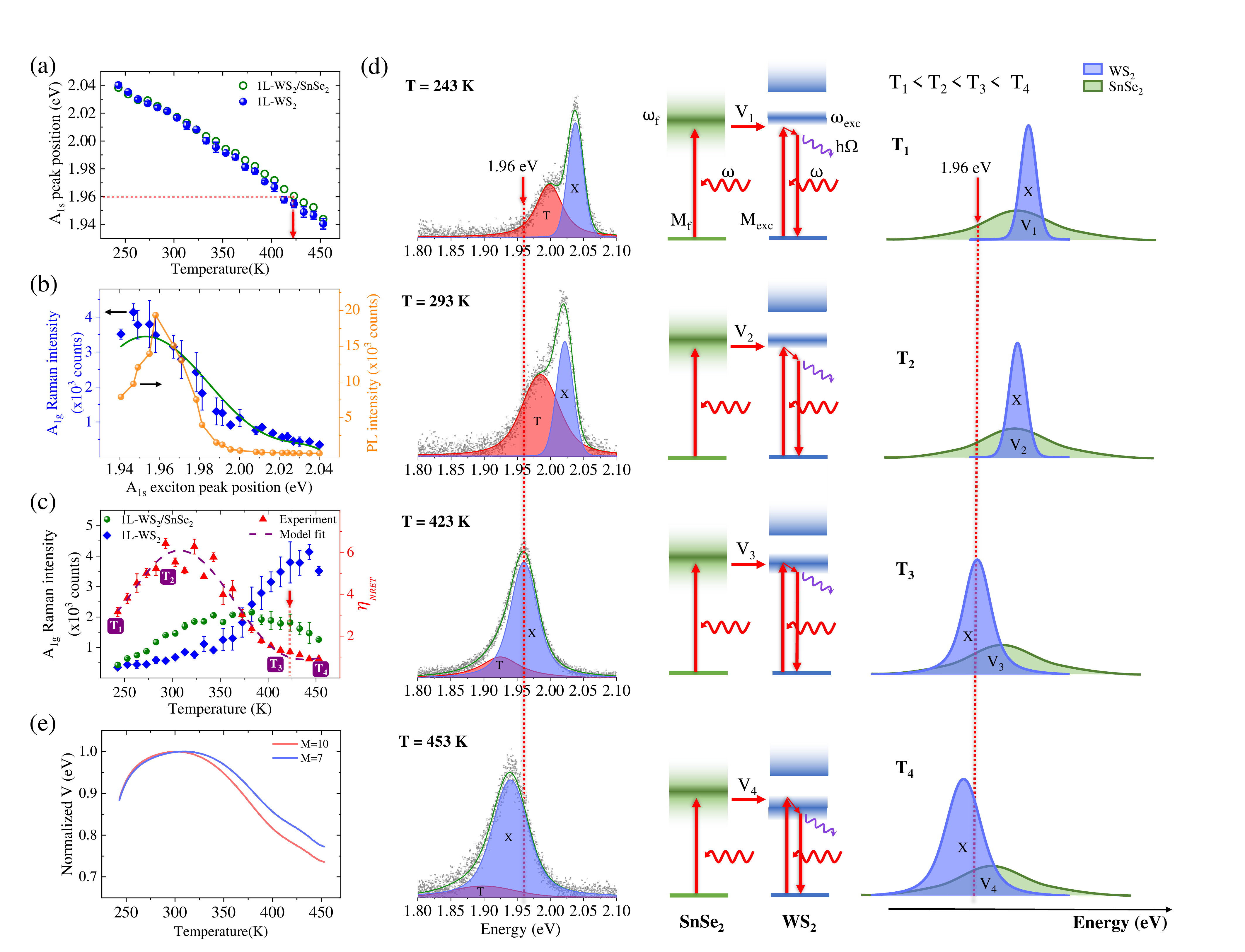}
	% \vspace{-0.1in}
	\caption{\textbf{Modulation of NRET Raman enhancement by spectral overlap tuning from temperature variation}. (a) Temperature versus $A_{1s}$ exciton position of WS\tsub2 on isolated WS\tsub2 (blue) and WS\tsub2/SnSe\tsub2 junction (green). (b) $A_{1s}$ exciton position dependent $A_{1g}$ Raman intensity profile (left axis) of isolated WS\tsub2 with $633$ nm excitation. Solid line represents corresponding fitting of the profile with model in equation \ref{eq:model1} that highlights excitation resonance. The corresponding PL intensity (with $633$ nm excitation) variation is shown on the right axis to emphasize excitation resonance with exciton. (c) Temperature dependent WS\tsub2 $A_{1g}$ Raman profiles (left axis) of isolated WS\tsub2 and WS\tsub2/SnSe\tsub2 junction. Right axis highlights the temperature modulated $\eta^{}_{_{NRET}}$ from the experiment (red symbols) and the model fit (dashed line) from equation \ref{eq:R}. (d) Left panel: PL spectra of WS\tsub2 from WS\tsub2/SnSe\tsub2 junction at different temperatures with deconvoluted exciton (X) and trion (T) peaks. The vertical dashed line shows the position of $633$ nm ($1.96$ eV) excitation. Middle panel: Schematic illustration of the NRET driven Raman scattering model from equation \ref{eq:R} with varying NRET strength ($V$) at different temperatures. These temperature points are highlighted in (c). Right panel: Schematic tunability of spectral overlap across WS\tsub2 and SnSe\tsub2 that modulates $V$ with temperature variation. (e) Extracted values of $V$ from the model fit of $\eta^{}_{_{NRET}}$ denoted in (c).}\label{fig:F4}
\end{figure*}
\newpage
\newpage


\section*{Supplementary material}

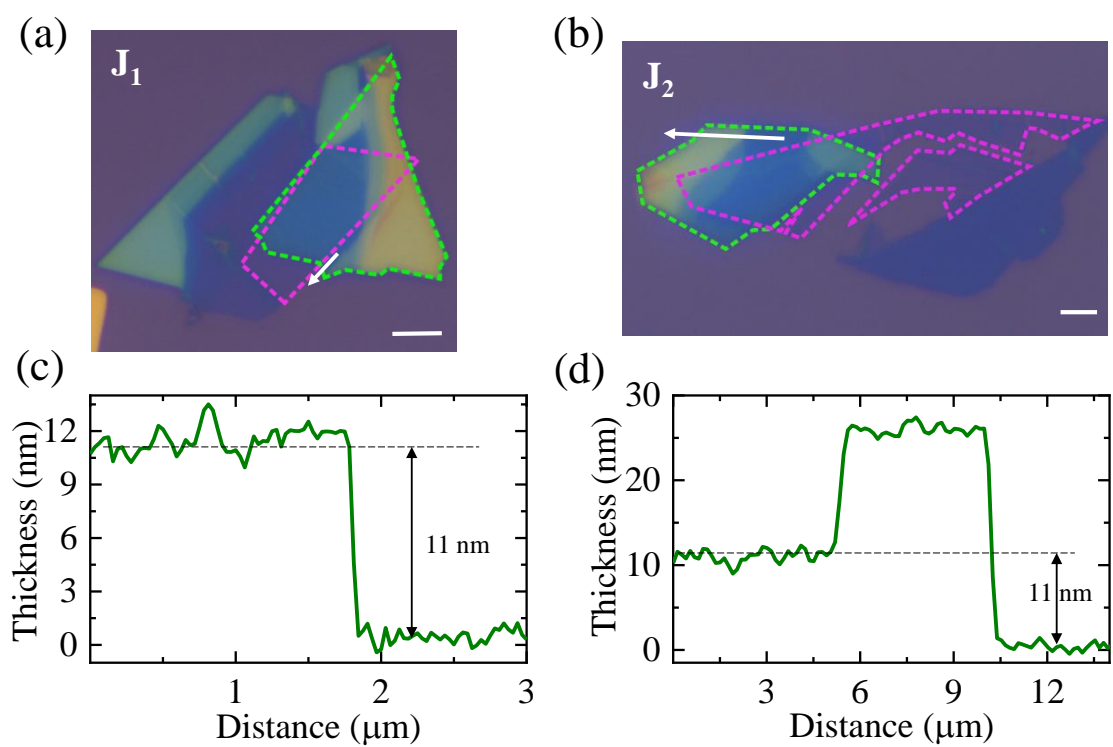

Figure S1: **Optical images and AFM of samples $J_1$ and $J_2$**. (a,b) Optical images of $WS_2/SnSe_2$ sample, $J_1$ and $MoS_2/SnSe_2$ sample, $J_2$. Dashed lines highlight 1L-TMD (pink) and $SnSe_2$ (green) regions. Scale bar is 5 μm. (c,d) Step height profiles of $SnSe_2$ along the white arrows in (a,b) obtained from AFM scans.

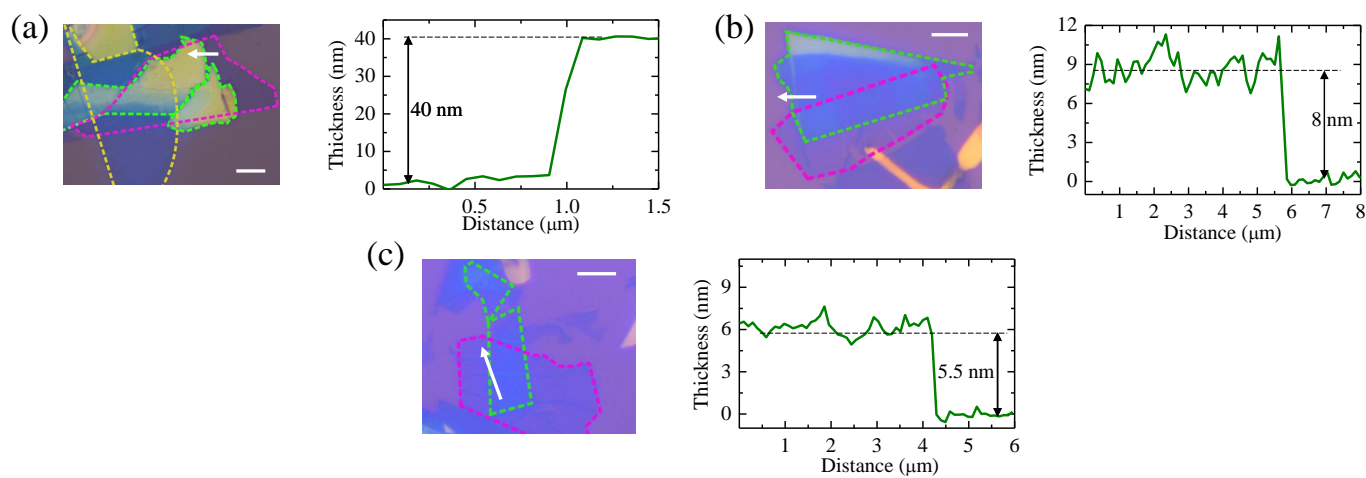

Figure S2: **Optical images and AFM of $MoS_2/SnSe_2$ samples**. Optical images of $MoS_2/SnSe_2$ samples and corresponding AFM step height profiles along the white arrows in optical images depicting the thickness of $SnSe_2$. Scale bar is 5 μm.

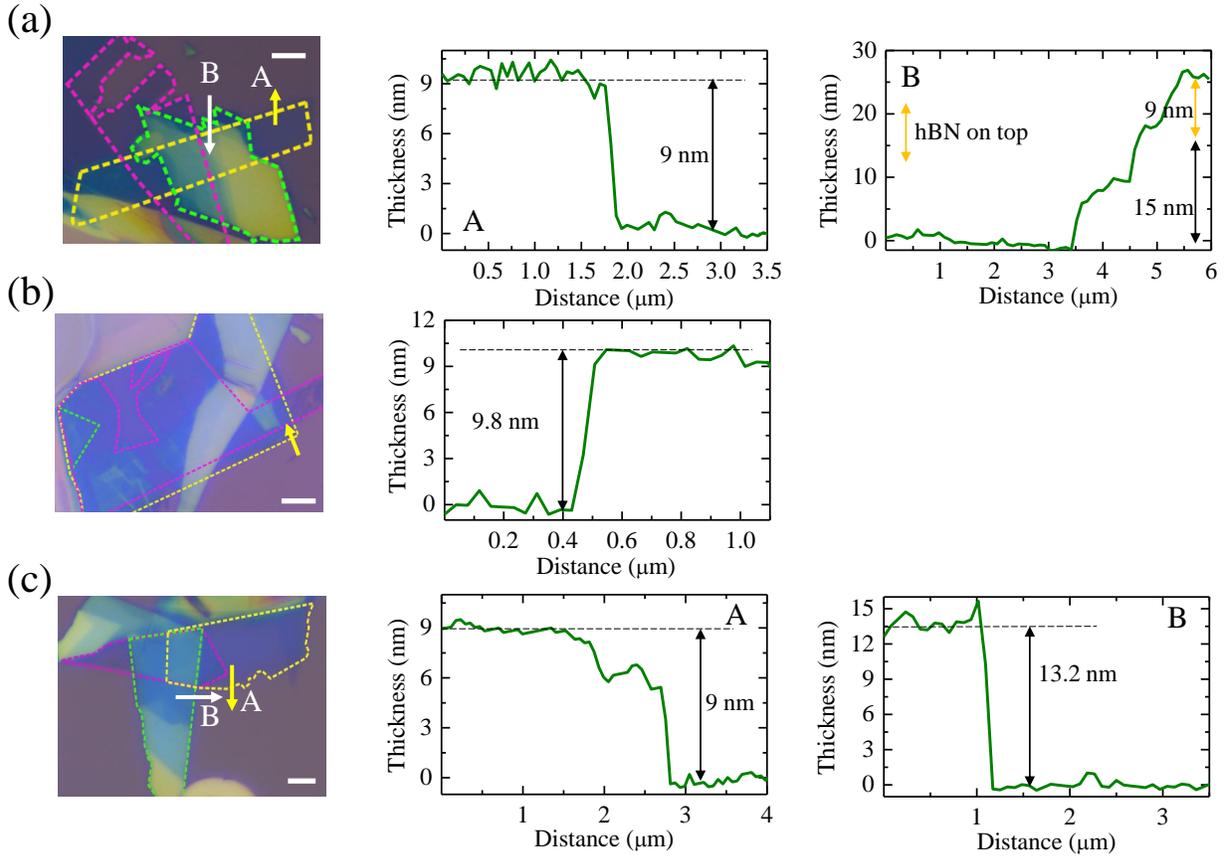

**Figure S3: Optical images and AFM of MoS$_2$/hBN/SnSe$_2$ samples**. Left panel shows the optical images of MoS$_2$/hBN/SnSe$_2$ samples. Scale bar is 5 $\mu$m. Dashed lines highlight the regions of MoS$_2$ (pink), hBN (yellow) and SnSe$_2$ (green) regions. Middle panel shows the corresponding AFM step height profiles of hBN along the yellow arrow in optical images. Right panel shows the corresponding AFM step height profiles along the white arrow in optical images from which SnSe$_2$ thickness is extracted.

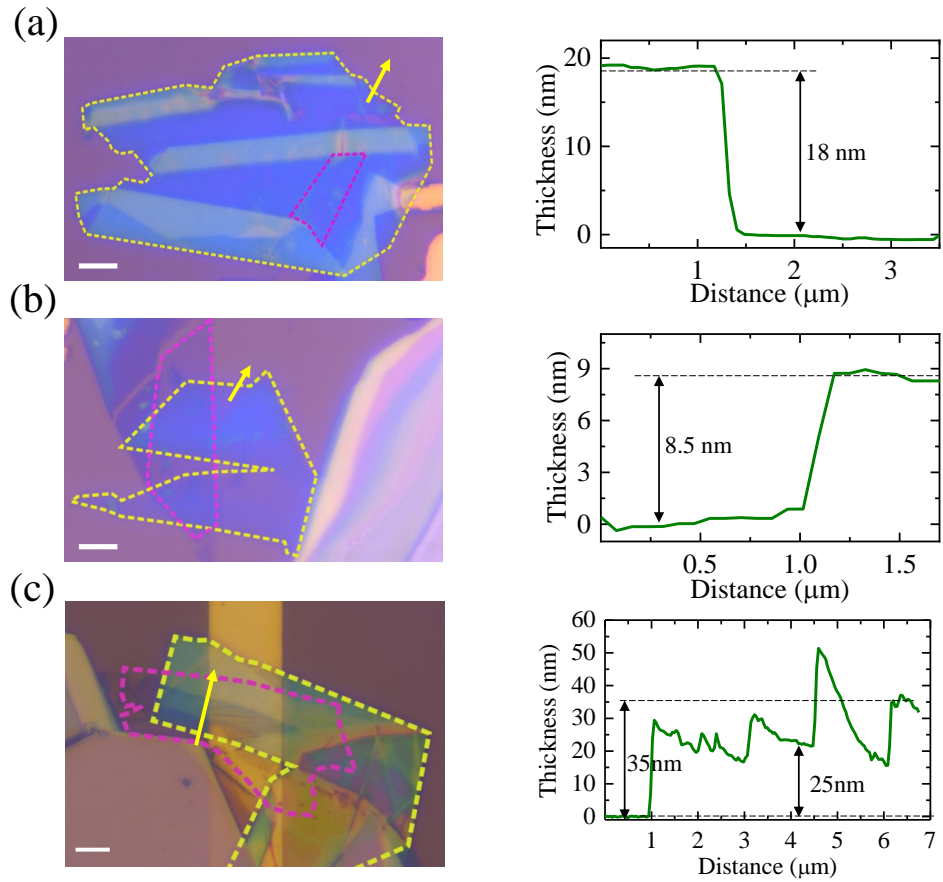

**Figure S4: Optical images and AFM of MoS$_2$/hBN samples.** Optical images of MoS$_2$/hBN samples (left panel) with the corresponding AFM step height profiles of hBN (right panel) along the yellow arrow in optical images. Scale bar is 5 $\mu$m. Region of hBN (MoS$_2$) is marked with yellow (pink) dashed line in optical images.

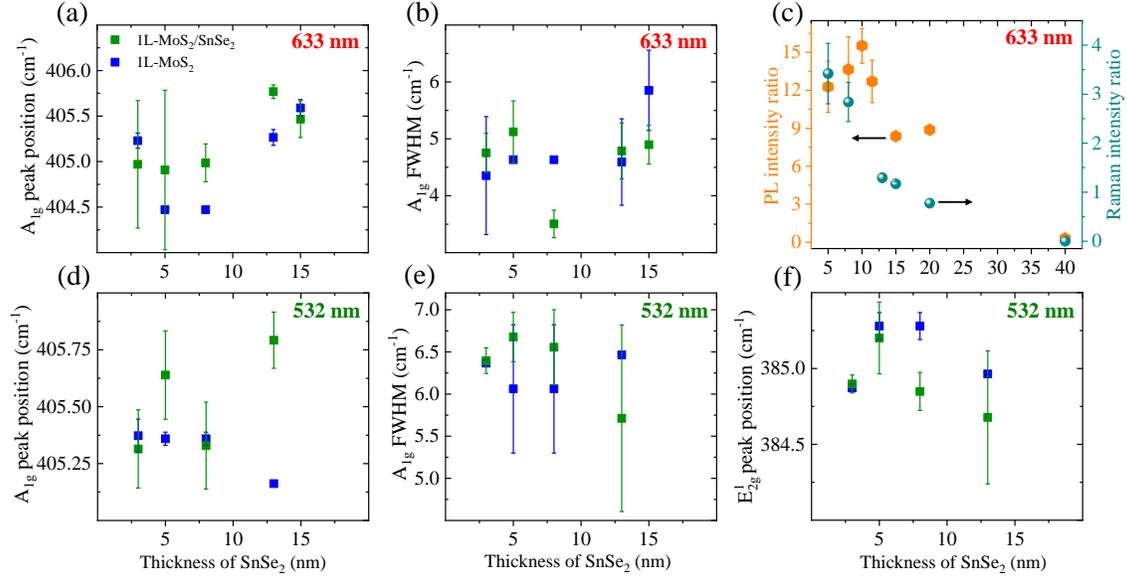

**Figure S5: Raman characterization from MoS$_2$/SnSe$_2$ samples**. (a,b) MoS$_2$ $A_{1g}$ peak position and FWHM with 633 nm excitation from isolated (blue) and junction (green) regions. (c) PL intensity ratio versus Raman intensity ratio across different MoS$_2$/SnSe$_2$ samples under 633 nm excitation emphasizing contribution of NRET and difference in their enhancement factors. (d,e) Peak position and FWHM of MoS$_2$ $A_{1g}$ with 532 nm excitation. (f) MoS$_2$ $E_{2g}^1$ peak position from isolated and junction regions from 532 nm Raman spectra.

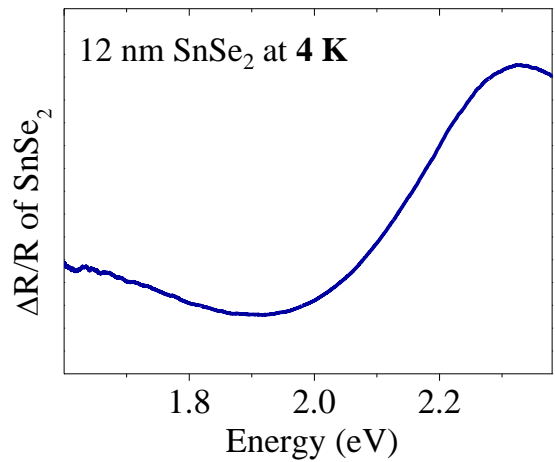

**Figure S6: Differential reflectance spectroscopy on SnSe$_2$**. Broad $\frac{\Delta R}{R}$ spectra of 12 nm SnSe$_2$ from differential reflectance spectroscopy at 4 K.

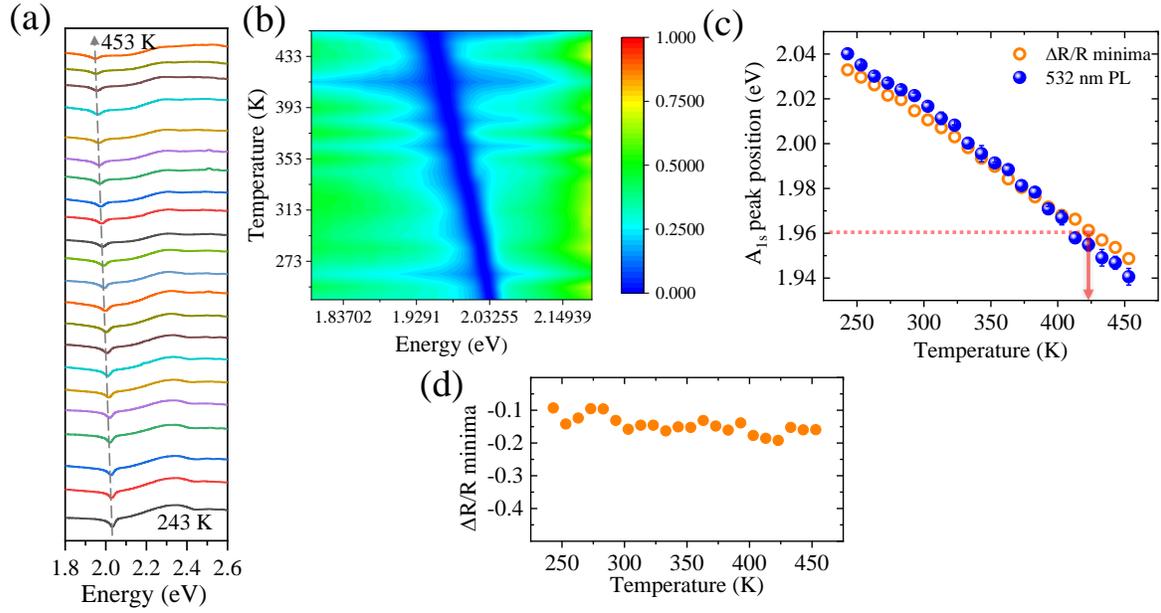

**Figure S7: Temperature dependent Differential reflectance spectroscopy on $WS_2$ in sample $J_1$.** (a) $\frac{\Delta R}{R}$ spectra from isolated $WS_2$ at different temperatures from 243 K to 453 K which represent shift of $A_{1s}$ exciton peak. (b) Temperature versus $A_{1s}$ peak energy contour plot from $WS_2$ $\frac{\Delta R}{R}$ spectra where blue region depicts the $\frac{\Delta R}{R}$ minima. (c) $A_{1s}$ peak position as a function of temperature extracted from $\frac{\Delta R}{R}$ minima and 532 nm PL. (d) Strength of $\frac{\Delta R}{R}$ minima as a function of temperature which shows relative similar oscillator strength of $WS_2$ from 243 K to 453 K.

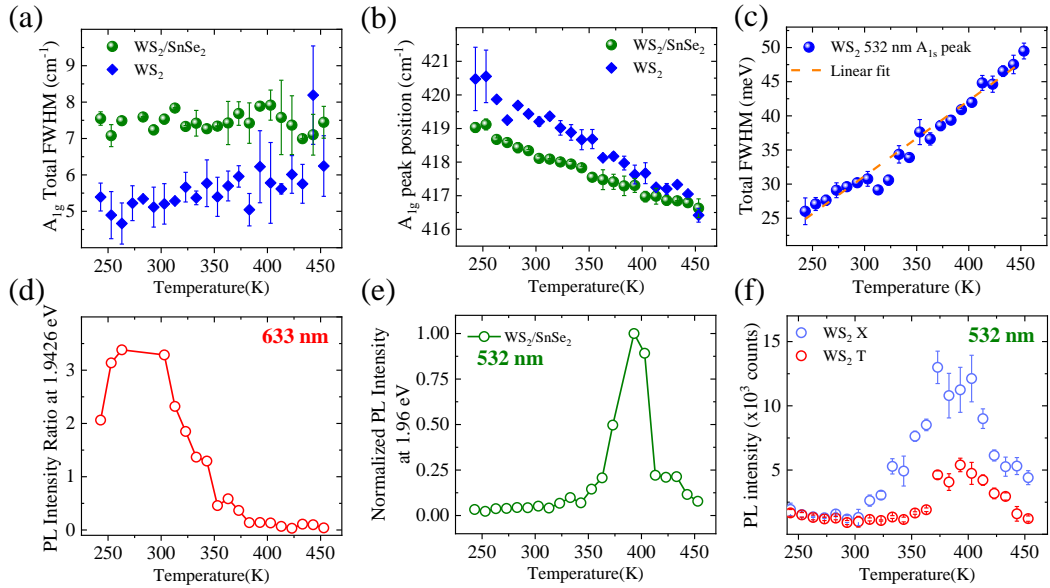

Figure S8: **Temperature dependent Raman and PL characterization of WS$_2$/SnSe$_2$ sample, J$_1$**. (a,b) Temperature versus WS$_2$ $A_{1g}$ FWHM and position on isolated (blue) and junction (green) regions from 633 nm excitation. (c) FWHM of WS$_2$ $A_{1s}$ exciton as a function of temperature and the corresponding linear fit. (d) 633 nm PL intensity ratio (at 1.9426 eV) of WS$_2$/SnSe$_2$ which exhibits modulation with temperature similar to $\eta_{NRET}$ discussed in the main text. (e) Normalized 532 nm PL intensity at 1.96 eV exhibiting maximum at the temperature close to 633 nm excitation resonance. (f) Exciton (X) and Trion (T) peak intensities as a function of temperature with 532 nm excitation.

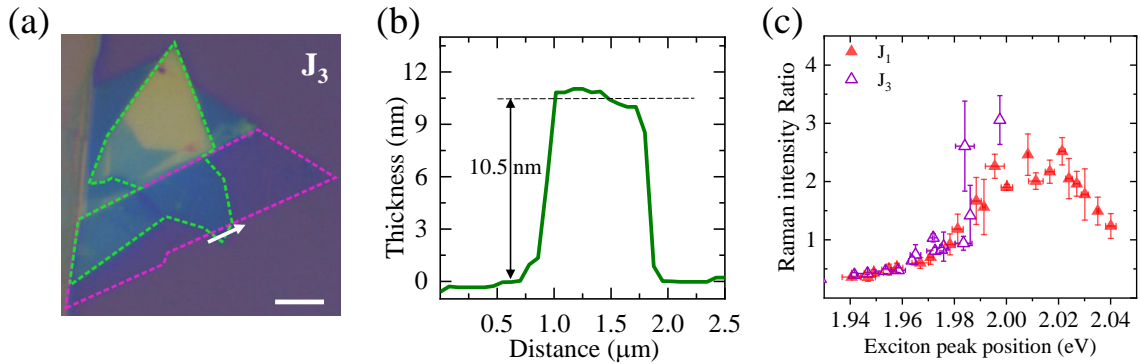

Figure S9: **Raman enhancement characteristics of another WS$_2$/SnSe$_2$ sample, J$_3$**. (a) Optical image of J$_3$ with WS$_2$ and SnSe$_2$ marked by pink and green dashed lines respectively. Scale bar is 5 $\mu$m. (b) SnSe$_2$ thickness profile from AFM along the white arrow in (a). (c) 633 nm WS$_2$ Raman intensity ratio as a function of exciton peak position from samples J$_1$ and J$_3$ which exhibit a similar trend.

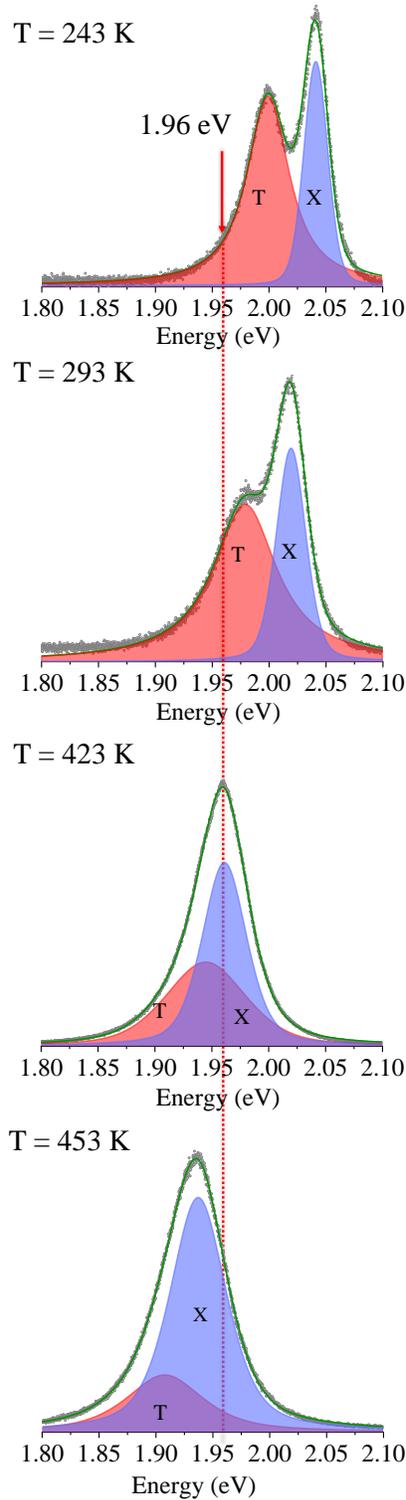

**Figure S10: Temperature dependent PL spectra of isolated WS$_2$ from WS$_2$/SnSe$_2$ sample, J$_1$.** PL spectra obtained from 532 nm excitation of isolated WS$_2$ at four different temperatures with corresponding fitting of exciton (X) and trion (T) peaks. Vertical dashed line indicates the position of 633 nm (1.96 eV) excitation.

# Graphical TOC Entry

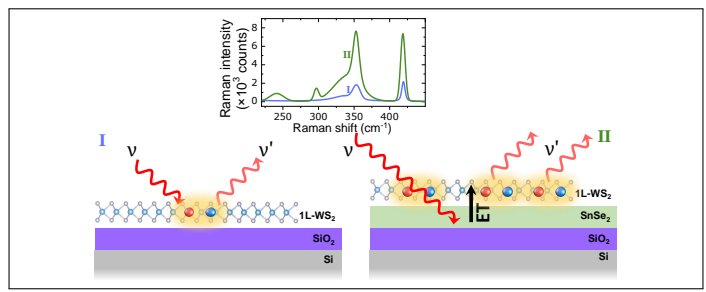


\end{document}